\documentclass[12pt]{article}
\pdfoutput=1
\usepackage{natbib}
 \bibpunct{(}{)}{;}{a}{,}{,}
\usepackage{graphicx}
\usepackage{amsmath,amssymb,amsthm}
\usepackage{mathtools}
\usepackage{mhequ}
\usepackage{bm}
\usepackage{rotating}
\usepackage[lined,ruled]{algorithm2e}
\usepackage{multirow}
\usepackage{rotating}
\usepackage{multicol}
\usepackage{titlesec}
\usepackage{latexsym}
\usepackage[pdftex,colorlinks=true,linkcolor=blue,citecolor=blue,urlcolor=blue,bookmarks=false,pdfpagemode=None]{hyperref}
\usepackage{url}
\usepackage{verbatim}
\usepackage{fancyhdr}
\usepackage{setspace}
\usepackage{paralist}
\usepackage{lineno}
\usepackage[top=1in, bottom=1in, left=1in, right=1in]{geometry}
\usepackage{lscape}
\usepackage{dcolumn}

\newcolumntype{.}{D{.}{.}{-1}}

\usepackage{enumerate} 
\usepackage{dsfont}

\def \be{\begin{equs}}
\def \ee{\end{equs}}
\newtheorem{theorem}{Theorem}[section]

\newtheorem{assumptions}[theorem]{Assumptions}

\newtheorem{example}[theorem]{Example}
\def \be{\begin{equs}}
\def \ee{\end{equs}}
\def \P{\mathbb{P}}
\def \E{\mathbb{E}}

\newcommand \gaussian{\mathrm{N}}

\begin{document}
\pagestyle{empty}


\title{Parallel Markov Chain Monte Carlo via Spectral Clustering}

\author{
 Guillaume W. Basse$^\dagger$ \and Natesh Pillai$^\ddagger$, \and Aaron Smith
 \thanks{
Guillaume Basse is a graduate student in the Department of Statistics at 
Harvard University (\href{mailto:gbasse@fas.harvard.edu}{gbasse@fas.harvard.edu})
 and his work was supported by a Google Fellowship in Statistics for 
North America. Natesh Pillai is an Associate Professor at Harvard University. Aaron Smith
is an Assistant Professor at the University of Ottawa
}}

\date{}

\maketitle
\thispagestyle{empty}

\newpage
\begin{abstract}
As it has become common to use many computer cores in routine applications, finding good ways to parallelize popular algorithms has become increasingly important. In this paper, we present a parallelization scheme for Markov chain Monte Carlo (MCMC) methods based on spectral clustering of the underlying state space, generalizing earlier work on parallelization of MCMC methods by state space partitioning. We show empirically that this approach speeds up MCMC sampling for multimodal distributions and that it can be usefully applied in greater generality than several related algorithms. Our algorithm converges under reasonable conditions to an `optimal' MCMC algorithm. We also show that our approach can be asymptotically far more efficient than naive parallelization, even in situations such as completely flat target distributions where no unique optimal algorithm exists. Finally, we combine theoretical and empirical bounds to provide practical guidance on the choice of tuning parameters.

\vfill
\end{abstract}

\newpage
\singlespacing
\small
\tableofcontents
\normalsize
\singlespacing

\newpage
\pagestyle{plain}
\setcounter{page}{1}


\section{INTRODUCTION}

Markov chain Monte Carlo (MCMC) is a powerful and popular method for sampling from target distributions. As a sampling method, it is inherently parallel: simply run independent copies of the Markov chain on every available core. However, as MCMC has been used for a wider variety of problems, it has become clear that this `naive' paralellization can often be improved upon. A major problem in the field is to develop new parallelization methods and find conditions under which they are better than the naive parallelization. 

Past approaches to parallelizing MCMC algorithms can be broadly divided into four categories: (i) \textit{single-step} speedups, (ii) \textit{exploration} speedups, (iii) learning or partitioning \textit{data} and (iv) learning or partitioning the \textit{state-space}. The first method uses several cores to increase the speed at which individual steps of a Metropolis-Hastings algorithm are taken (\cite{wilkinson05, calderhead14, brockwell06, angelino14, feng03}), while the second method uses several cores to run slightly different chains, increasing the speed at which the algorithm mixes (see \textit{e.g.}, parallel-tempering in \cite{altekar04} and the elliptical slice sampler in \cite{nishihara14}). Contrary to the first two approaches, the third approach applies mainly in the context of sampling from posterior distributions, and involves partitioning the data into batches and later recombining the MCMC samples (\cite{scott13, wang13,neiswanger13, huang2005}). This approach is especially useful for large data sets as they are often stored across different machines. The fourth method often involves finding good partitions of the state space and running a different MCMC chain in each part of the partition (\cite{hallgren14, vanderwerken13}), though there exist other methods in this category (\cite{CRY09}). Although we describe four categories, these methods can generally all be applied at the same time. In addition, several data-augmentation chains incorporate the underlying dataset into the state space of the Markov chain (see, e.g., \cite{maclaurin2014firefly}), meaning that statespace-partitioning schemes, including the approach described in this paper, can be used as a first step in data-partitioning schemes.\\

\subsection{Our Contributions}

We propose a novel collection of methods for parallelizing MCMC by partitioning the underlying state space. Our key idea is to find partitions based on spectral clustering (see \cite{vonluxburg07}). The main intuition behind all state space partitioning methods is to replace a single Markov chain targeting a highly multimodal distribution with several Markov chains, each targeting distinct unimodal distributions. Since Markov chains tend to mix much more quickly on unimodal distributions than on multimodal distributions, this should improve computational efficiency. Our approach differs from existing space-partitioning approaches in that it includes a general way to find a partition (unlike \cite{hallgren14}) and that the allowable partitions are extremely general and can in particular include non convex sets (this is in contrast to the more limited family of Voronoi partitions suggested in \cite{vanderwerken13}). Perhaps surprisingly, we find that the additional flexibility of our family of partitions generally does not greatly increase the cost of finding a `good' partition, \textit{as long as a `good' partition exists}. \\
We provide empirical evidence that our approach works well for benchmark problems and provide examples for which our method outperforms its competitors. We also provide a theoretical grounding for our approach. This includes some guarantees that the partitions we find converge to a `good' partition and that `good' partitions result in efficient MCMC chains. We use fully worked-out examples to illustrate the gains that our method can provide under optimal circumstances, as well as the fact that state space partitioning can give an advantage over naive parallelization even when the target distribution does not have strong clusters and when the partitions used are neither stable nor close to optimal. Finally, we discuss heuristics for the amount of computational effort that should be spent on finding a partition.

\section{Intuition Behind Partitioning}
\label{section:intuition}

Before discussing how our algorithm chooses a partition, we give notation and explain why state space partitioning methods work well once a good partition has been found.  Let $\{Y_{t} \}_{t \in \mathbb{N}}$ be a Markov chain on a state space $\Omega$ with stationary distribution $\pi$ and transition kernel $K$. Throughout this paper, we assume that the kernel $K$ is a Metropolis-Hastings kernel associated with a proposal kernel $Q$, though our approach can be applied in other settings.  For any $T \in \mathbb{N}$ and $\pi$-measurable function $h$, the usual MCMC estimate of $\mu \equiv \pi(h)$ is 
\be \label{EqStandardEst}
\hat{\mu} = \frac{1}{T} \sum_{t=1}^{T} h(Y_{t}).
\ee 
The computation of this estimate can be naively parallelized by running $n$ independent chains $\{ Y_{t}^{(i)} \}_{t \in \mathbb{N}, 1 \leq i \leq n}$ and writing
\be \label{EqNaiveParEst}
\hat{\mu}_{\mathrm{naive}} = \frac{1}{nT} \sum_{t=1}^{T} \sum_{i=1}^{n} h(Y_{t}^{(i)}).
\ee 
Alternatively, fix a partition of $\Omega$ into disjoint subsets $\{ \Omega_i \}_{i=1}^{n}$. Define the weights $w_{i} = \pi(\Omega_{i})$ and distributions 
\be \label{EqRestrictedStatDists} 
\tilde{\pi}_i(A) = \pi(A \cap \Omega_i)  \mbox{ and } \pi_{i}(A) = \frac{1}{w_{i}} \tilde{\pi}_i(A).
\ee
For $1 \leq i \leq n$, define $K_{i}$ to be the Metropolis-Hastings kernel with proposal kernel $Q$ and target distribution $\pi_{i}$ and let $\{ X_{t}^{(i)} \}_{t \in \mathbb{N}}$ be a Markov chain evolving according to $K_{i}$. For each fixed $i$, the estimate of $\mu_{i} \equiv \pi_{i}(h)$ that is analogous to \eqref{EqStandardEst} is given by
\be \label{EqDefMeanEstPartOnly}
\hat{\mu}_{i, \mathrm{par}} = \frac{1}{T} \sum_{t=1}^{T} h(X_{t}^{(i)}),
\ee 
and since $\mu = \sum_{i} w_{i} \mu_{i}$, we use
\be \label{EqPartParEst}
\hat{\mu}_{\mathrm{par}} = \sum_{i=1}^{n} w_{i} \hat{\mu}_{i, \mathrm{par}} 
\ee 
as the full estimate of $\mu$. \\
We now derive conditions under which the estimator in \eqref{EqPartParEst} has smaller variance than the estimator in \eqref{EqRestrictedStatDists}. Let $ \lambda = \sup \{ | \lambda^* | \, : \, (\lambda^* I - K)^{-1} \text{ is not a bounded linear operator on } L^{2}(\pi), \, \lambda^* \neq 1 \}$ and denote by $(1-\lambda)$ the spectral gap of reversible kernel $K$, and similarly by $(1-\lambda_i)$ the spectral gap of $K_i$.
The normalized variance of the estimate \eqref{EqNaiveParEst} can be bounded (see e.g. Prop 4.29 of \cite{aldous-fill-2014}) by
\be \label{IneqVarFormNaive}
T \, \mathrm{Var}[\hat{\mu}_{\mathrm{naive}}](1 - O(T^{-1})) \leq \nu_{\mathrm{naive}} \equiv \frac{2 || h ||_{2, \pi}^{2}}{n(1 - \lambda)}, 
\ee 
and generically there exists a function $h$ for which this is close to equality for large $T$. Similarly,
\be \label{IneqVarFormPar}
T \, \mathrm{Var}[\hat{\mu}_{\mathrm{par}}](1 - O(T^{-1})) \leq \nu_{\mathrm{par}} \equiv 2  \sum_{i}  \frac{w_{i}^{2} || h ||_{2, \pi_{i}}^{2}}{1 - \lambda_{i}},\,\,\,\,\,\,\,
\ee 
and again this is generically close to equality for large $T$ and worst-case $h$. Denote by $\Phi$ the set of all measurable $n$-partitions of $\Omega$ and by $\mathcal{P} = (\Omega_{1}, \ldots, \Omega_{n})$ an element of $\Phi$. Equations \eqref{IneqVarFormNaive} and \eqref{IneqVarFormPar} suggest that the estimate \eqref{EqPartParEst} obtained from a partitioned state space should be more efficient than the naive estimate \eqref{EqNaiveParEst} when
\be \label{IneqParNecCond}
\sum_{i} \frac{n w_{i}^{2} (1  - \lambda)}{1 - \lambda_{i}} \frac{||h||_{2,\pi_{i}}^{2}}{||h||_{2,\pi}^{2}} < 1.
\ee 
Since $||h||_{2,\pi_{i}}^{2} = \int_{x} h(x)^{2} \pi_{i}(dx) \leq \frac{1}{w_i}|| h ||_{2,\pi}^{2}$ for all $i$, this suggests choosing the partition
\be \label{EqRealObjectiveFunction}
\mathcal{P} = \mathrm{argmin}_{\mathcal{P} \in \Phi} \left( \sum_{i} \frac{w_{i} }{1 - \lambda_{i}} \right)
\ee 
to find an estimator that is efficient for generic functions $h$. Finding this partition is computationally difficult, and so we settle for finding a partition that makes a good proxy for $\sum_{i} \frac{w_{i}}{1 - \lambda_{i}}$ small. Define the \textit{conductance} of $K_{i}$ by
\be \label{eq:conductance}
\phi_{i}(S) &= \frac{\int_{x \in S} K_{i}(x, S^{c}) d \pi_{i}(x)}{\pi_{i}(S) \pi_{i}(S^{c})} \\
\phi_{i} &= \inf_{S \, : \, 0 < \pi_{i}(S) < \frac{1}{2}} \phi_{i}(S),
\ee 
with an analogous definition for the conductances associated with $K$. By \cite{lawler88}, 
\be \label{eq:conductanceVgap}
\frac{\phi_{i}^{2}}{2} \leq (1 - \lambda_{i}) \leq 2 \phi_{i},
\ee 
and so 
\be \label{IneqCondCons}
\sum_{i} \frac{w_{i} }{1 - \lambda_{i}} \leq 2 \sum_{i} \frac{w_{i}}{\phi_{i}^2}.
\ee 
Thus, we approximately minimize the objective function \eqref{EqRealObjectiveFunction} by making the upper bound \eqref{IneqCondCons}  small. It is known (see \cite{meila00, kannan04}) that spectral clustering approximately finds 
\be \label{EqSpecClust}
\mathcal{P} = \mathrm{argmin}_{\mathcal{P} \in \Phi} \sum_{i=1}^{n} w_{i} \phi(\mathcal{P}_{i}).
\ee 
Although it is not obvious, this choice of partition also approximately minimizes the right-hand side of Equation \eqref{IneqCondCons} (see \cite{lee2014multiway}), and so throughout this paper we will generally choose our partitions via spectral clustering.  By inequalities \eqref{IneqVarFormNaive}, \eqref{IneqVarFormPar} and \eqref{IneqCondCons}, the condition 
\be \label{IneqBeatNaiveCond}
\sum_{i} \frac{n w_{i}^{2} (1  - \lambda)}{\phi_{i}^{2}} \frac{||h||_{2,\pi_{i}}^{2}}{||h||_{2,\pi}^{2}} \ll 1
\ee 
implies that $\nu_{\mathrm{par}} \ll \nu_{\mathrm{naive}}$. \\

Inequality \eqref{IneqBeatNaiveCond} gives a sufficient condition under which partitioning results in a more efficient sampler than naive parallelization. We give some examples showing that the `optimal' partition defined by the heuristic \eqref{EqSpecClust} can satisfy this condition, and also that even very poor approximations of this partition can vastly improve sampling efficiency. The following illustrates the enormous improvement that partitioning can achieve when each mode of a strongly multimodal target density is in a separate part of the partition: \\

\begin{example}[Mixture of Gaussians]
Fix constants $0 < \tau \ll \sigma \ll 1$ and consider the Random Walk Metropolis-Hastings chain $K$ on $\mathbb{R}$ with proposal kernel $Q(x,\cdot) = \frac{1}{2 \tau} \mathrm{Unif}[x-\tau, x+ \tau]$ and target distribution $\pi = \frac{1}{2} \gaussian(-1, \sigma) + \frac{1}{2} \gaussian(1,\sigma)$. By considering the set $S = (-\infty, 0]$, we can calculate from Equation~\eqref{eq:conductance} that this chain has conductance (and thus spectral gap by Equation \eqref{eq:conductanceVgap}) at most $O \left( \tau^{-2} e^{- c \sigma^{-2} } \right)$ for some fixed $ 0 < c < \infty$. We consider speeding up simulation from the target distribution by partitioning the state space $\mathbb{R}$ into $n=2$ parts. Although spectral clustering attempts to minimize the objective function \eqref{EqSpecClust} rather than the `correct' objective function \eqref{EqRealObjectiveFunction}, Figure ~\ref{fig:ObjFunctComp} suggests that both have the same minimizer: the partition $\mathcal{P} = \{ (-\infty,0], (0,\infty) \}$. 

\begin{figure}[h] 
\begin{center}
\includegraphics[width=0.5\textwidth]{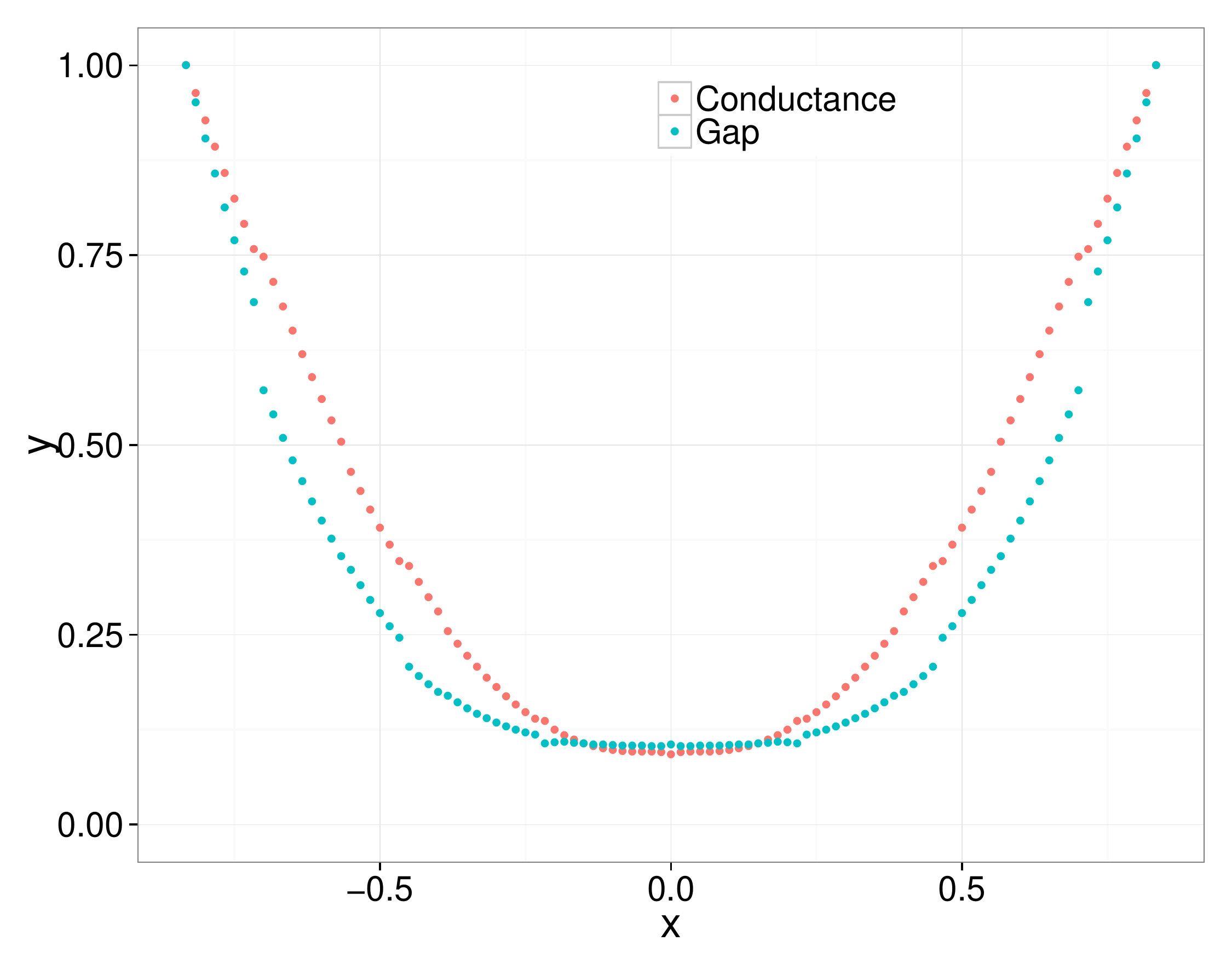}
\caption{Comparing Two Objective Functions ($\sigma = 0.4$, $\tau = 0.2$) (Red is objective function for spectral clustering; blue is objective function for MCMC).}
\label{fig:ObjFunctComp}
\end{center}
\end{figure}

The Metropolis-Hastings chains $K_{1}$ and $K_{2}$ with proposal distribution $Q$ and target distributions $\pi_{1}(x) \propto \pi(x) \textbf{1}_{x \leq 0}$ and $\pi_{2}(x) \propto \pi(x) \textbf{1}_{x > 0}$ have conductances that are at least on the order of $\frac{\tau}{\sigma}$ in the same regime (see Theorem 4.3.3 of \cite{woodard07}), and thus spectral gaps that are at least on the order of $\frac{\tau^{2}}{\sigma^{2}}$ (again, see Equation \eqref{eq:conductanceVgap}). Thus, the ratio of efficiencies \eqref{IneqBeatNaiveCond} is at most $O(\frac{1}{\sigma^{2}} e^{- c \sigma^{-2} }) \ll 1$. Figure ~\ref{fig:EffRat} plots the the inverse of the true value of this ratio, showing that it is enormous for reasonable values of $\sigma$ and illustrating the gains of state space partitioning over naive parallelization. Note - this plot is on a logarithmic scale!

\begin{figure}[h] 
\begin{center}
\includegraphics[width=0.45\textwidth]{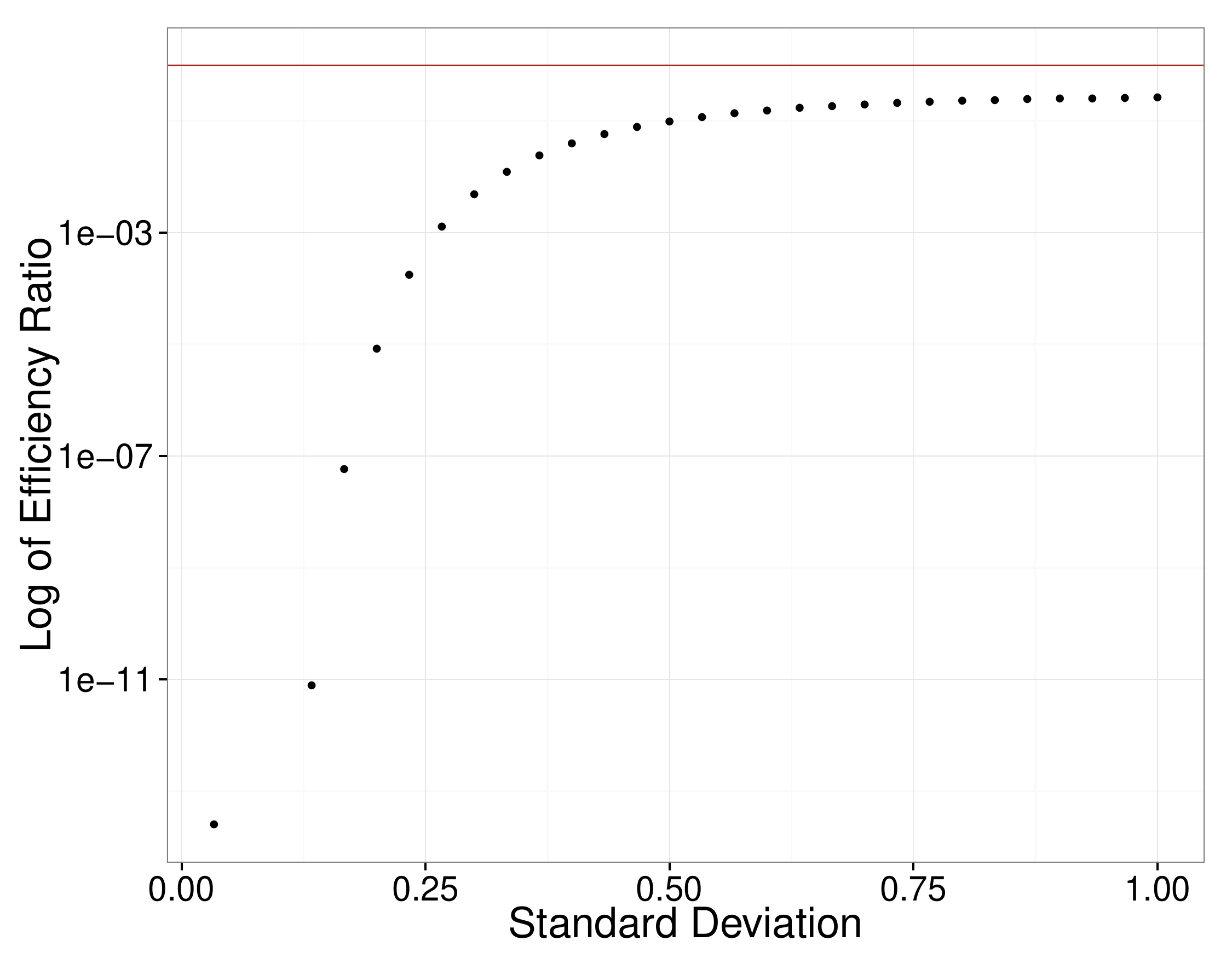}
\caption{Log of Efficiency Ratio ($log(Var(\hat{\mu}_{par}) / Var(\hat{\mu}_{naive})$ for $\tau = 0.2$).}
\label{fig:EffRat}
\end{center}
\end{figure}

Figure ~\ref{fig:ObjFunctComp} suggests that spectral clustering is essentially optimizing the right objective function, and Figure~\ref{fig:EffRat} shows that the estimator $\hat{\mu}_{\mathrm{par}}$ associated with the best partition can be vastly more efficient than $\hat{\mu}_{\mathrm{naive}}$. Unfortunately, in realistic examples, we will not have access to an optimal partition. Thus, it is natural to ask if similar gains can be obtained for partitions that are closer to those that might be seen in practice. Figure ~\ref{fig:EffDec} shows the relative efficiency of $\hat{\mu}_{\mathrm{par}}$ as the partition $\mathcal{P} = \{ (-\infty, R], (R,\infty) \}$ ranges over various values of $R \geq 0$. It shows that $\hat{\mu}_{\mathrm{par}}$ can be much more efficient than $\hat{\mu}_{\mathrm{naive}}$ even when the partition used is quite far from optimal. 

\begin{figure}[h] 
\begin{center}
\includegraphics[width=0.45\textwidth]{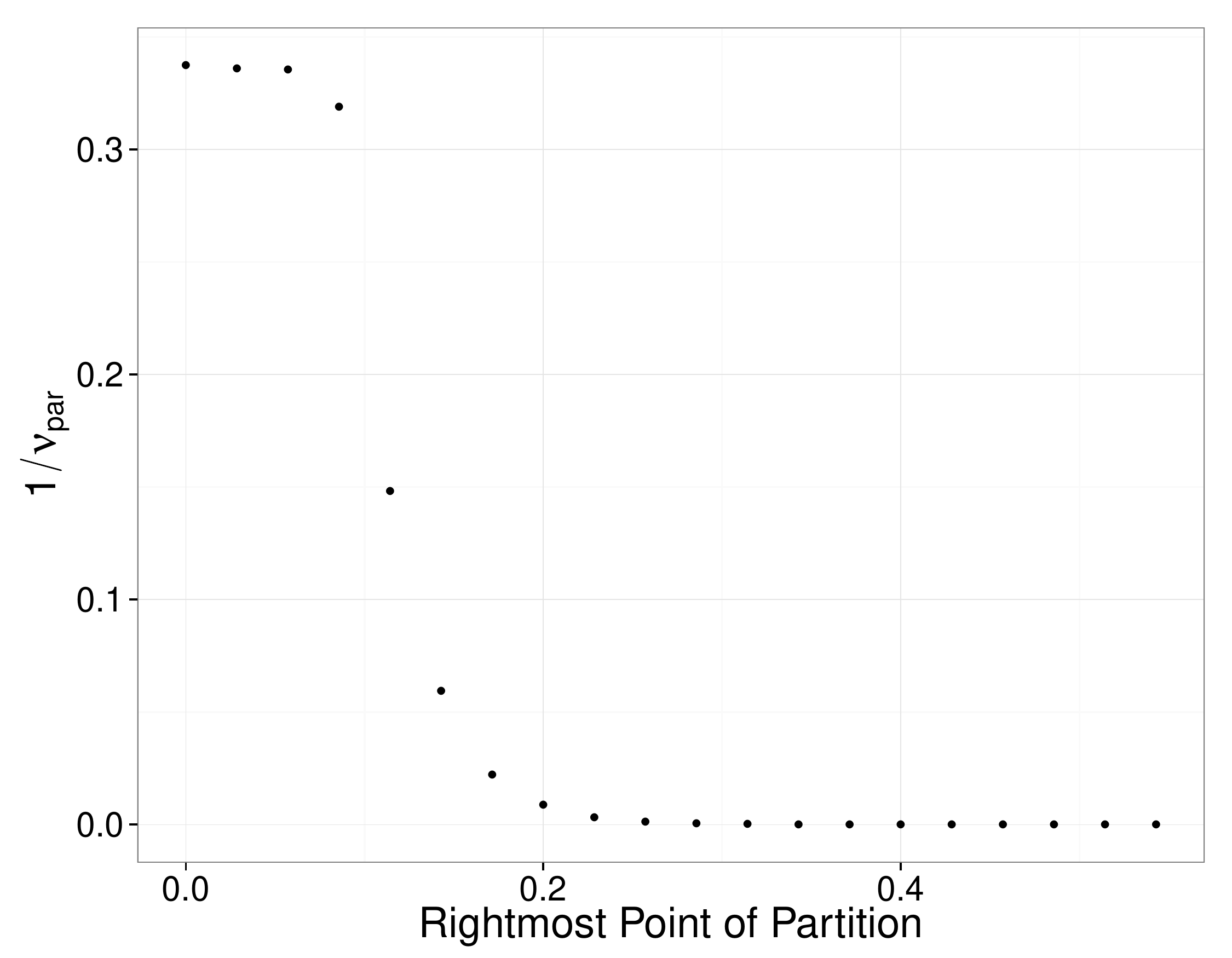}
\caption{Inverse of the bound of the variance of $\hat{\mu}_{\mathrm{par}}$ as $R$ changes ($\tau = 0.2$, $\sigma = 0.15$).}
\label{fig:EffDec}
\end{center}
\end{figure}

Together, the three Figures  ~\ref{fig:ObjFunctComp}, ~\ref{fig:EffRat} and ~\ref{fig:EffDec} justify our approach by showing that the partition returned by spectral clustering is closely related to the best partition, that the best partition can substantially increase efficiency, and finally that there is a fairly wide range of partitions that give rise to estimators with nearly-optimal efficiency. 
\end{example}

A central point of this paper is that our parallelization scheme can offer large improvements even when the partition used is far from optimal, and even in the absence of multimodality. This is most starkly illustrated by: \\

\begin{example}[Simple Random Walk on the Cycle] \label{ExSrwCycle}
Fix $m$ and define the graph $(V_{m}, E_{m})$ with $V_{m} = \{1,2,\ldots, m\}$ and $E_{m} = \{ (x,y) \in V_{m} \, : \, | x - y | \leq 1 \} \cup \{ (1,m) \}$. Recall that the simple random walk on the circle (here $\Omega = V_{m}$), which has transition kernel $\P[X_{t+1} = y | X_{t} = x] = \frac{1}{3} \textbf{1}_{ (x,y) \in E_{m}}$, has stationary distribution $\pi = \mathrm{Unif}(\Omega)$ and spectral gap of order $\frac{1}{m^{2}}$ (see Section 12.3 of \cite{levin08}). Again, we fix a function $h$ with $|| h ||_{2,\pi}^{2} = 1$ and compare the efficiency of $\hat{\mu}_{\mathrm{par}}$ to the efficiency of $\hat{\mu}_{\mathrm{naive}}$. By Equation \eqref{IneqVarFormNaive}, the variance of $\hat{\mu}_{\mathrm{naive}}$ is on the order of $\frac{m^{2}}{n T}$ . If we partition $\Omega$ into $n$ connected components $\{ \mathcal{P}_{j} \}_{j=1}^{n}$, the associated kernels $K_{j}$ have spectral gaps on the order of $\frac{1}{| \mathcal{P}_{j} |^{2}}$ (again, see \cite{levin08}). Thus, by Equation \eqref{IneqVarFormPar}, the associated estimator $\hat{\mu}_{\mathrm{par}}$ has variance on the order of $\frac{1}{m^{2}T} \sum_{j=1}^{n} | \mathcal{P}_{j} |^{4}$. It is easy to check that this is minimized by choosing $| \mathcal{P}_{j} | = \frac{m}{n}$, in which case $\hat{\mu}$ has an asymptotic variance of only $\frac{m^{2}}{n^{3} T}$. This is much smaller than the variance $\frac{m^{2}}{n T}$ of $\hat{\mu}_{\mathrm{naive}}$, despite the fact that the target distribution is completely flat.  \\
We note again that suboptimal partitions can yield substantial improvements. For example, any choice of partition for which $\sup_{j} | \mathcal{P}_{j} | = o \left( \frac{m}{\sqrt{n}} \right)$ will lead to an estimator $\hat{\mu}_{\mathrm{par}}$ with a smaller asymptotic variance than the estimator $\hat{\mu}_{\mathrm{naive}}$. We also note that, due to the rotational symmetry of the simple random walk, there is not a unique optimal partition in this example. This non-uniqueness does not prevent us from using the algorithm presented in this paper. In most clustering applications, the resulting partition is of interest in and of itself, and is meaningful only if a unique good partition exists. In contrast, we only use the partitioning method as a way to reduce the variance of our estimator, and don't attach any particular meaning to it: we only need the existence of at least one good partition.
\end{example}

The remainder of this paper is concerned with describing an algorithm that produces a good partition efficiently, as well as describing its performance.

\section{Methods}

In this section, we lay out our approach and give some useful variations. Our main approach, summarized in Algorithm 1, has four steps: an initial exploration step, followed by repeated partitioning, sampling and weighting. In the first step, we explore the state space and try to capture as many modes as possible; this is necessary if we hope to have a reasonable first partition.  In the second step, we use spectral clustering and the history of the algorithm to find a `good' partition $\{ \Omega_{i} \}_{i=1}^{n}$ of the state space - that is, one for which the restricted chains $K_{i}$ all have large conductance. In the third step, we run the chains $\{X_{t}^{(i)} \}_{t \in \mathbb{N}}$ in each component $\Omega_{i}$ of the partition. In the last step, we estimate the weights $w_{i}$ of each element of the partition. The algorithm requires as input the number $n$ of cores to be used (corresponding to the number of disjoint sets in the partitions), the proposal kernel $Q$, the target distribution $\pi$, the number $N_{0}$ of samples to be obtained by the initial exploration stage, the number $\ell$ of times that repartitioning occurs, the number $\{ N_{i} \}_{1 \leq i \leq \ell}$ of samples used in each repartitioning step (first step of Algorithm~\ref{algo:spectralclust}), and the number $\{T_{i} \}_{1 \leq i \leq \ell}$ of steps to run the Markov chain between repartitionings. In practice, it is often helpful to have $\ell$, $T$ and especially $N$ depend on the previous history of the chain rather than fixing them in advance.

\begin{algorithm}
	\SetKwFunction{partitionFn}{partitionFn}
	\SetKwFunction{DoSpectralClustering}{DoSpectralClustering}
	\SetKwFunction{ExploreStateSpace}{ExploreStateSpace}
	\SetKwFunction{runParallelChains}{RunParallelChains}
	\SetKwFunction{unnormalizedRestrictedDensity}{unnormalizedRestrictedDensity}
	\SetKwInOut{Input}{input}\SetKwInOut{Output}{output}
	\Input{Q, $\pi$,n, $N_{0}$, $\ell$,  $\{N_{i}\}_{1 \leq i \leq \ell }$, $\{T_{i}\}_{1 \leq i \leq \ell }$, }
	\Output{$\hat{\mu}$} \
	\Begin{
	Initialize $X$ as in Section \ref{SubsecExpl}\;
	\For{$ i = 1$ \KwTo $\ell$}{
		$(\Omega_{1}, \ldots, \Omega_{n}), (V_{1}, \ldots, V_{n}) \leftarrow$ \DoSpectralClustering($X, n, N_{i}, Q, \pi$)\;
		Compute $(\tilde{\pi}_1, ..., \tilde{\pi}_n)$ as in Equation~\eqref{EqRestrictedStatDists}  \;
		$X \leftarrow X \cup$ \runParallelChains($Q, \tilde{\pi}_{1}, \tilde{\pi}_{2}, \ldots, \tilde{\pi}_{n}, T_{i}$)\;
		Compute $(\hat{w_1}, ..., \hat{w_n})$ as in Equations~\eqref{eq:bridge1} and \eqref{eq:bridge2} \;
	}
	Estimate $\hat{\mu}$ as in Equations~\eqref{eq:estimate2}\;
	}
	\caption{Our Method}
	\label{algo:method}
\end{algorithm}

\subsection{Explore State Space} \label{SubsecExpl}

In this step, we create a sample $X= (X_{1}, X_{2}, \ldots, X_{N_{0}})$ of points from the state space $\Omega$; these points will be used to create an initial partition. Ideally, these points should cover every mode of the target $\pi$. In simulations, we have found that generating these points by running the parallel tempering algorithm (\citet{swendsen86, geyer91,earl05}) from several well-dispersed initial points works well. 

\subsection{Partition State Space} \label{SubsecPart}

In the \textit{partitioning step} we obtain a partition of the state space, given the collection of points $X$ that we have seen so far. We summarize our approach in Algorithm 2. 

\begin{algorithm}
	\SetKwFunction{cluster}{cluster}
	\SetKwInOut{Input}{input}\SetKwInOut{Output}{output}
	\SetKwFunction{kmeans}{kmeans}
	\Input{X, n, N, Q, $\pi$}
	\Output{$(\Omega_1, ..., \Omega_n)$, $(V^1, .., V^n)$}
	\Begin{
	Subsample N points uniformly and without replacement $X_{1}, \ldots, X_{N} \sim \mathrm{Unif}(X)$ \;
	Define the matrix $\widehat{Q}_{ij} = Q(X_i, X_j)$ for $i, j \in \{1, ..., N\}$ \;
	Define the diagonal matrix $D_{ii} = \sum_j \widehat{Q}_{ij}$ for $i \in \{1, ..., N\}$\;
	Let $L = D^{-1/2} \widehat{Q} D^{-1/2}$\;
	Let $V^1, ..., V^n$ be the $n$ normalized leading eigenvectors of $L$\;
	For i=1,...,N let $Z_i = (V^{1}[i], ..., V^{n}[i]) / \|(V^{1}[i], ..., V^{n}[i])\|$\;
	Define the map $\sigma : \, \{Z_{1},Z_{2}, \ldots, Z_{N} \} \rightarrow \{ 1,2, \ldots, n\}$ by \kmeans($Z_{1}, \ldots, Z_{N}; n$) \;
	Let $\{C_1, \ldots, C_n\}$ be the $n$ centers obtained by \kmeans\;
	Define the partition $\Omega = \sqcup_{i=1}^{n} \Omega_i$ by Equation \eqref{EqPartitionExtension}\;
	}
	\caption{DoSpectralClustering}
	\label{algo:spectralclust}
\end{algorithm}

 \medskip

The first step, in which we subsample $N$ points from the original data $X$, is used to  keep the computational burden manageable, as the dataset can be very large. The second-last step of this algorithm, \textbf{kmeans}, is the popular $k$-means clustering algorithm (see \textit{e.g.}, Chapter 13 of \cite{HTF13}). The last step is to extend the partition $\sigma$ of the set $\{ Z_{1}, \ldots, Z_{N} \}$ to a partition of the entire state space $\Omega$. Let $\lambda_{1}, \ldots, \lambda_{n}$ be the eigenvalues associated with $V^{1}, \ldots, V^{n}$. Following Equations (8) through (12) of \cite{bengio03}, for $x \in \Omega$ and $1 \leq i \leq n$, we define
\be \label{EqPointExtension}
Z_{i}(x) = \frac{\sqrt{N}}{\lambda_{i}} \sum_{j=1}^{N} V^{i}[j] Q(x,x_{j}).
\ee
Set $Z(x) = (Z_{1}(x), \ldots, Z_{n}(x))$ and for $1 \leq i \leq n$ define
\be \label{EqPartitionExtension}
\Omega_{i} = \{ x \in \Omega \, : \, \underset{j \in \{1, \ldots, n\}}{\mathrm{argmin}}( ||C_j - Z(x)||) = i\}. \,\,\,\,\,\,\,\,
\ee
\subsection{Run Chains} \label{SubsecRun}

We define the method \textbf{RunParallelChains}. For each $i \in \{1,2,\ldots, n\}$, we let $\{ X_{t}^{(i)} \}_{t = 1}^{T}$ be a Metropolis-Hastings chain with proposal kernel $Q$ and target distribution $\tilde{\pi}_{i}$. The method then returns $\{ X_{t}^{(i)} \}_{1 \leq t \leq T, \, 1 \leq i \leq n}$. We do not specify the initial points $X_{1}^{(i)}$, but have found in practice that the point in $X \cap \Omega_{i}$ corresponding to the centroid given by the $k$-means algorithm is a good choice.

\subsection{Estimate Weights} \label{SubsecCompWeights}

The final step is to estimate the weights $w_{i} = \pi(\Omega_{i})$. There are a few options here, but we find that bridge sampling (\cite{Meng96},\cite{gelman98}) works well. Recall that bridge sampling requires, for each $1 \leq i \leq n$, a proposal distribution $p_i$ and bridge function $\alpha_{i}$. In this paper, we generally choose $p_{i}$ to be a normal or student distribution whose first two moments match the empirical moments of $\mathrm{Unif}(X_{i})$ and use the geometric bridge $\alpha_i = (p_i \tilde{\pi}_i)^{-1/2}$, where $\tilde{\pi}_i (x) = w_i \pi_i(x)$ is the unnormalized version of $\pi_i$.

For fixed $p_{i}$ and $\alpha_{i}$, let $X_{i} = X \cap \Omega_{i}$, $n_{i} = | X_{i} |$, and let $\{ \theta_{j} \}_{1 \leq j \leq n_{i}}$ be i.i.d. draws from $p_{i}$. Then, define
\begin{equation} \label{eq:bridge1}
	\hat{c}^{(i)}_1 = \frac{\frac{1}{n_{i}} \sum_{j=1}^{n_{i}} \tilde{\pi}_i(\theta_{j})\alpha(\theta_{j})}{\frac{1}{n_{i}} \sum_{\theta \in X_{i}} p_i(\theta)\alpha(\theta)}.
\end{equation}
Since the estimates $\hat{c}^{(1)}_1, ..., \hat{c}^{(n)}_1$ do not generally add up to 1, we renormalize:
\begin{equation} \label{eq:bridge2}
	\hat{w}_i = \frac{\hat{c}^{(i)}_1}{\sum_{j=1}^n \hat{c}^{(j)}_1}.
\end{equation} 
\subsection{Estimating $\hat{\mu}$}

We conclude by defining an estimator for $\hat{\mu}$. We start by writing: 
\begin{equation}\label{eq:muhatest}
	\mu = \mathbb{E}_{\pi}(h(x)) = \sum_{i=1}^n \mu_i w_i
\end{equation}
\noindent where $\mu_i = \pi(\Omega_i)\mathbb{E}_{\pi}(h(x) 1_{x\in\Omega_i}) = \mathbb{E}_{\pi_i}(h(x))$. We estimate $\mu_i$ by $\hat{\mu}_i$ as in Equation \eqref{EqDefMeanEstPartOnly}, and using the estimated weights $\hat{w}_1, .., \hat{w}_n$ we have the following estimator for $\mu$:
\begin{equation} \label{eq:estimate2}
	\hat{\mu} = \sum_{i=1}^n \hat{\mu}_i \hat{w}_i
\end{equation}

\section{Applications}
\label{section:applications}

We describe the parameters and proposals used for the simulations in the Supplementary Material. 

\subsection{Example 1: Mixture of Gaussians in 2 dimensions}

\begin{figure}[h]
	\centerline{\includegraphics[width=.4\textwidth]{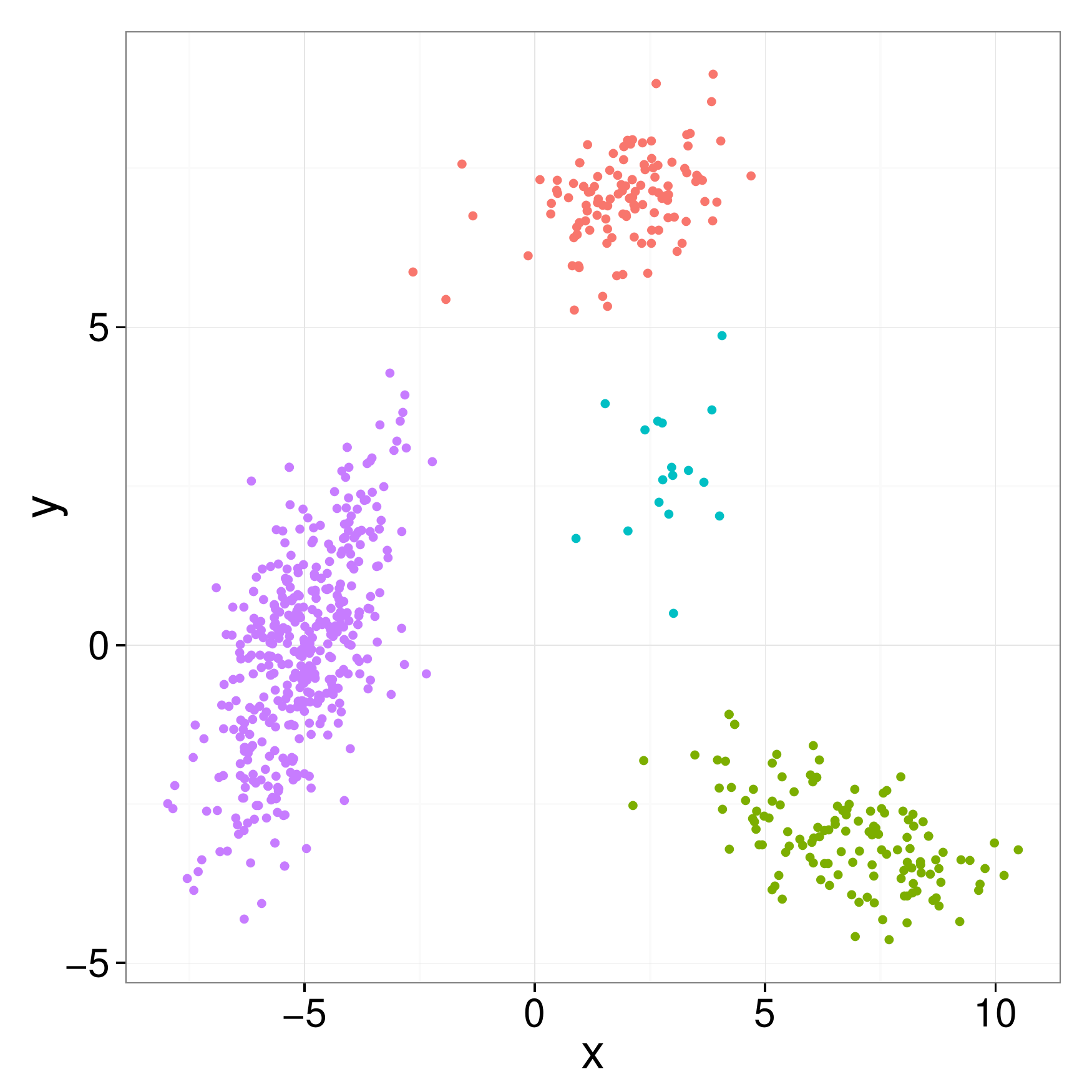}}
	\caption{Spectral clustering of sample space for 2D mixture of gaussians}\label{fig:mixture2d}
\end{figure}

We start with an example used by \cite{vanderwerken13}, where the target distribution here is a mixture of bivariate normals. We compare the performance of three methods: parallel tempering, naive parallelization, and our method. For our method, we obtained $N_0 = 8000$ samples using parallel tempering as our exploration phase (see Figure~\ref{fig:mixture2d}). We then ran our algorithm with $\ell = 1$ round of partitioning, $n=4$ clusters, and $N_1 = 700$, thus obtaining estimates $\hat{w}_i, i=1, \ldots, 4$ of the weights. Considering those weights as fixed, we ran parallel constrained chains for an additional $T_{1} = 4000$ iterations, and obtained an estimate $\hat{\mu}$ as in \eqref{eq:muhatest}.

To evaluate our method we repeated the last step (the last 4000 iterations) 500 times, computed the Euclidian distance between our estimate and the true expectation for each replication, and computed the average squared error, and the sample standard deviation of the squared error. In total, for one estimate, we ran a total $8000 + 4000 + 4 \times 4000 =  28000$ iterations. But if we assume that each iteration (whether it be for parallel tempering or constrained metropolis) takes t seconds, and consider the fact that $4\times 4000$ iterations are run in parallel on 4 cores, each estimate is obtained in $16000\times t$ seconds. For parallel tempering, we obtained one estimate of the mean by running the algorithm for 16000 iterations. We again repeated this 500 times to obtain an average squared error and a sample standard deviation. For the naive method, we ran in parallel 4 independent chains initialized randomly from the target distribution, for 16000 iterations. In the end, if we consider that an iteration of parallel tempering takes the same time than an iteration of metropolis hastings, then all the methods take the $16000 \times t$ seconds (in fact, one iteration of parallel tempering takes a bit longer; thus, as measured by clock time, our method would look even better). 

\begin{table}[h]
	\centering
	\begin{tabular}{c|c|c}
	\hline
	method & mean & se\\
	\hline
	ours & 0.008 &  0.009\\
	parallel tempering & 0.21 & 0.28\\
	naive parallel & 14.08 & 13.5\\
	\hline
	\end{tabular}
	\caption{Square distance from the true mean in the 2D mixture of gaussians example}
	\label{tab:vanderwerken2d}
\end{table}

We see in Table~\ref{tab:vanderwerken2d} that our method dramatically reduces the mean squared error compared to parallel tempering, and improves on naive parallelization even more dramatically.

\subsection{Example 2: Why Spectral Clustering?} 
\begin{figure}[h!]
	\centerline{\includegraphics[width=.5\textwidth]{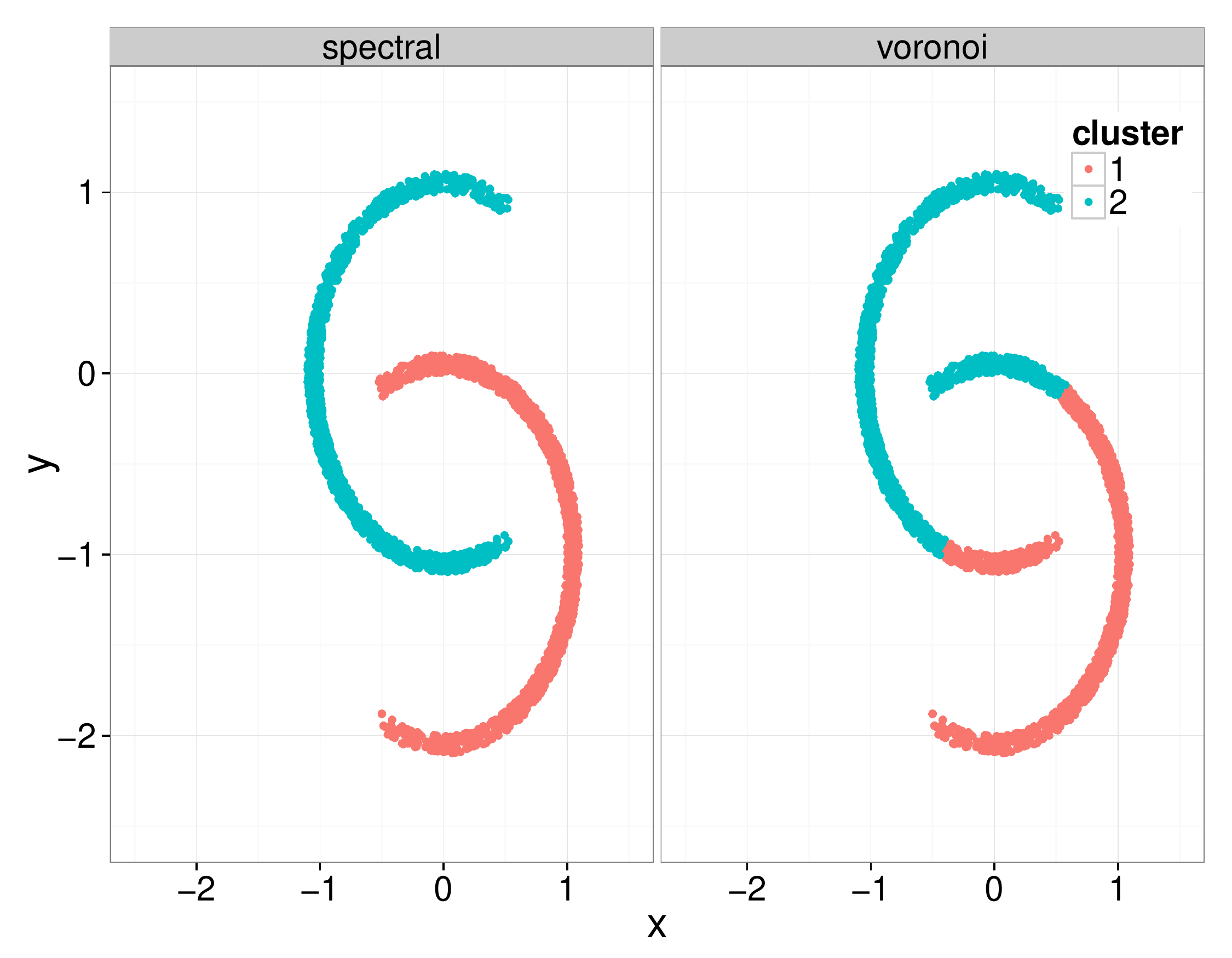}}
	\caption{Spectral (left) and k-means (right) clusterings of the sample space}
	\label{fig:sshape}
\end{figure}

%
In this section, we illustrate the flexibility of spectral clustering, and show that it can work in situations where Voronoi clustering fails. Define the two sets $\mathcal{S}_1 = \{ \left(r\cos(\theta), r\sin(\theta)\right) :  \theta\in [\frac{2\pi}{6}, \frac{10\pi}{6}], r\in [1, 1.1] \}$ and $\mathcal{S}_2 = \{ \left(r\cos(\theta), (r-1)\sin(\theta) \right) : \theta\in [\frac{-4\pi}{6}, \frac{4\pi}{6}], r\in [1, 1.1] \}$, and consider the target distribution $\pi$ that is uniform on $\mathcal{S} = \mathcal{S}_1 \cup \mathcal{S}_2$ (see Figure~\ref{fig:sshape}). We consider the simplified scenario where the exploration is carried by two chains of length 5000, initialized in the two sets, which results in a reasonably good picture of the overall density. We then perform spectral clustering on the one hand, and k-means (an instance of Voronoi clustering) on the other hand, using the $N_0 = 10000$ samples obtained in the exploration step. Figure~\ref{fig:sshape} show the results of the clustering phase: we see that because k-means can only find convex partitions, it doesn't capture the shapes adequately. To illustrate the impact of the choice of clustering method in terms of Monte-Carlo error, we carried the rest of our algorithm (estimating of the weights, running restricted parallel chains, and estimating the mean of the distribution) with parameters $(\ell = 1, N_1=700, T_1=10000)$ for each clustering method . In particular, we simulated the last two steps of algorithm 1 (running restricted chains and estimating the mean) 200 times for each method, and computed the squared Euclidian distance of the estimated mean to the true mean. Using spectral clustering, the average squared distance is 0.04, with standard error 0.05, while with k-means, the average distance is 0.13, with standard error 0.03.
%
\section{Convergence and Optimality} \label{section:theory}
%
All proofs are in the Supplementary Material.
\subsection{Consistency}

We do not assume that the sequence of partitions we obtain at each stage $1 \leq i \leq \ell$ of Algorithm \ref{algo:method} is in any sense optimal, or that it converges in any sense to a good partition. Indeed, as illustrated by example \ref{ExSrwCycle}, our clustering algorithm can greatly increase computational efficiency even when there is not a unique optimal partition and when the partition used is far from any optimal partition. Even when there is not convergence to a unique optimal partition, the estimator $\hat{\mu}$ of $\mu$ returned by Algorithm \ref{algo:method} is generally consistent. We consider a simple setting:

\begin{assumptions} \label{AssSimple}
Assume that the state space $\Omega \subset \mathbb{R}^{d}$ is bounded and that the target distribution $\pi$ and the proposal distributions $\{ Q(x,\cdot) \}_{x \in \Omega}$ have densities $\rho(\cdot)$ and $q(x,\cdot)$ that satisfy
\be \label{DensityAssumption}
c < \rho(y), q(x,y) < C
\ee 
for some $0< c < C < \infty$ and all $x,y \in \Omega$.
\end{assumptions}

\begin{theorem}
Let Assumptions \ref{AssSimple} hold, and assume further that $h$ is bounded. Fix $\ell$ and $\{T_{i}\}_{i=1}^{\ell - 1}$, and let $\hat{\mu}$ be the estimate returned by Algorithm \ref{algo:method}. Then
\be 
\P[\lim_{T_{\ell} \rightarrow \infty} \hat{\mu} = \mu] = 1.
\ee 
\end{theorem}

Although our method doesn't require that our partitions converge to an optimal partition, this convergence is desirable and does occur under reasonable conditions. For a partition $\mathcal{P} = \{ \Omega_{i} \}_{i=1}^{n}$ of $\Omega$, define an equivalence relation on $\Omega$ by writing $x \sim_{\mathcal{P}} y$ if and only if there exists some $1 \leq i \leq n$ such that $x,y \in \Omega_{i}$. Then define the distance $d_{\pi}$ between pairs of partitions $\mathcal{P} = \{ \Omega_{i} \}_{i=1}^{n}$,  $\mathcal{P}' = \{ \Omega_{i}' \}_{i=1}^{n}$ by 
\be 
d_{\pi}(\mathcal{P}, \mathcal{P}') = \P[(X \sim_{\mathcal{P}} Y) \oplus (X \sim_{\mathcal{P}'} Y)],
\ee 
where $X,Y$ are drawn independently from $\pi$ and $\oplus$ denotes the logical operator \textit{XOR}. For any kernel $Q$ and distribution $\pi$ on $\Omega$, \cite{VLBB08} defines an associated limiting Laplacian $\mathcal{L} = \mathcal{L}(Q,\pi)$. For any limiting Laplacian $\mathcal{L}(Q,\pi)$, \cite{BDVLP06} defines the notion of a class of partitions $\mathcal{C} = \mathcal{C}(Q,\pi)$ associated with $\mathcal{L}$. We do not give precise definitions of these objects in this paper; the only heuristic needed is that $\mathcal{C}(Q,\pi)$ generally has exactly one element, unless $Q,\pi$ have symmetries. We can then state the following corollary to Theorem 16 of \cite{BDVLP06}:

\begin{theorem} [Convergence of Partitions] \label{ThmConvPart}
Let Assumptions \ref{AssSimple} hold. Fix a partition $\mathcal{P} = \{\Omega_{j}\}_{j=1}^n$ with associated measures $\{ \pi_{i} \}_{i=1}^{n}$, so that  $\mathcal{C}(Q, \frac{1}{n} \sum_{j=1}^{n} \pi_{j})$ has a unique element $\mathcal{P}$. Fix $\gamma > 0$ and two sequences $\{ N(k) \}_{k \in \mathbb{N}}$, $ \{ T(k) \}_{k \in \mathbb{N}}$ satisfying 
\be \label{ConvClustAssumption}
\lim_{k \rightarrow \infty} T(k) &= \infty \\
\lim_{k \rightarrow \infty} \frac{N(k)^{2 + 2\gamma}}{T(k)} &= 0.
\ee 

For $k \in \mathbb{N}$, let $X = \{X_{t}^{(i)}\}_{0 \leq t \leq T(k), \, 1 \leq i \leq n}$ be the output of the method {\em\textbf{RunParallelChains}}$(Q,\pi_{1},\ldots,\pi_{n},T(k))$ in  Algorithm \ref{algo:method}. Let $P_{k,X}$ be the partition returned by  {\em \textbf{DoSpectralClustering}}$(X,n,N(k),Q,\pi)$. Then for $\epsilon > 0$,
\be \label{EqLimConv}
\lim_{k \rightarrow \infty} \P[d_{\pi}(\mathcal{P}_{k,X}, \mathcal{P} ) > \epsilon] = 0.
\ee 
\end{theorem}

This result implies that, if the partition $\{ \Omega_{i} \}_{i=1}^{n}$ at stage $1 \leq q < \ell$ of  Algorithm ~\ref{algo:method} is close to the optimal partition as measured by the metric $d_{\pi}$ on partitions, the metric at stage $q+1$ can be made arbitrarily close as well by choosing $N_{q+1}$, $T_{q+1}$ large.


\subsection{Sample Size Heuristics}

In this section, we discuss the choice of the sample size $N$ used to compute each partition in Algorithm \ref{algo:method}. We emphasize two facts:
\begin{enumerate}
\item Increasing $N$ has essentially no impact on the mixing properties of $K_{i}$ after a certain point $N_{\max}$.
\item If there exists an optimal partition $\{ \Omega_{i} \}_{i=1}^{n}$, and $d_{i,j} = \E_{x \sim \pi_{i}, y \sim \pi_{j}}[\|x - y \|]$ represents the distance between parts of the partition while $1 - \lambda_{i}$ is the spectral gap of kernel $K_{i}$, we often have 
\be \label{IneqHeuristic} 
N_{\max} \approx \left( \frac{\max_{1 \leq i < j \leq n} d_{ij}}{\min_{1 \leq i \leq n} \pi(\Omega_{i}) (1 - \lambda_{i})} \right)^{2}.
\ee 
\end{enumerate}

\begin{figure} 
\includegraphics[width=0.45\textwidth]{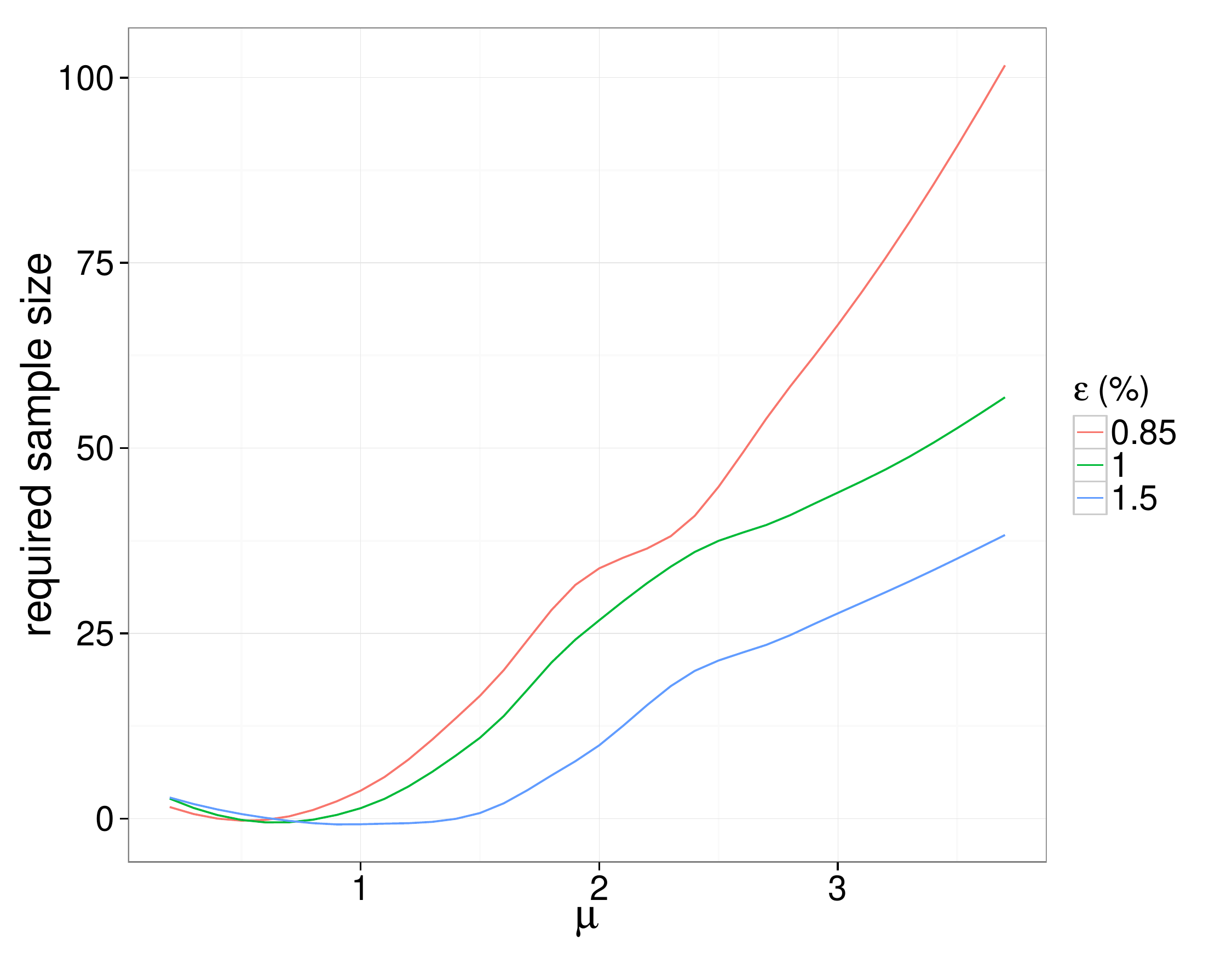}
\caption{Number of points needed to create a partition within a certain distance of the sample optimal partition}
\label{NMaxHeur}
\end{figure}

Together, these tell us that for the problems where our methods are most useful (\textit{i.e.} where $\max_{1 \leq i \leq n} (1-\lambda_{i})$ is largest), the amount of effort that spent on finding the partitions should be small. 

The first heuristic follows from the fact that the partitions returned by \textbf{DoSpectralClustering} converge to the optimal partition under moderate conditions (see Theorem \ref{ThmConvPart}) and that the mixing time and spectral gap are continuous functions of the underlying transition kernel (see \textit{e.g.} the main result of \cite{mitrophanov2005sensitivity}). 

Our justification for the second heuristic is empirical. Let $Q(x, x^*) = \frac{I(|x-x^*|<\tau)}{2\tau}$, and for $0 < \mu < \infty$ let $\pi_{\mu} = \frac{1}{2} \mathcal{N}(-\mu,1) + \frac{1}{2} \mathcal{N}(\mu,1)$. For any partition $\mathcal{P} = \{\Omega_{i}\}_{i=1}^{n}$, let $\lambda(\mathcal{P})$ be the smallest spectral gap of the associated kernels $\{ K_{i} \}_{i=1}^{n}$ and let $\mathcal{P}_{0} = \{ (-\infty, 0], [0,\infty)\}$. Finally, for $0 < \epsilon <1$, define
\be \label{IneqNMaxDef}
N_{\max}(\epsilon, \mu) = \min \{ N  :  \lambda(\mathcal{C}_{N,\mu}) \geq (1 - \epsilon) \lambda(\mathcal{C}_{0}) \},
\ee 
the number of points needed to create a partition that is within a factor of $1 - \epsilon$ of the optimal partition.

For each $\mu \in  \{0.2, 0.3, \ldots, 3.7\}$ we generated i.i.d. samples $X_{\mu} = \{X_{i}\} \sim \pi_{\mu}$ and generated partitions $\mathcal{P}_{N,\mu}$ for $N = \{1,2,\ldots\}$, according to \textbf{DoSpectralClustering}$(X_{\mu},2,N,Q,\pi_{\mu})$. Figure \ref{NMaxHeur} presents smoothed versions of the averages of the curves $\{ N_{\max}(\epsilon, \mu) \}_{\mu \in \{0.2, 0.3, \ldots, 3.7\}}$ for $\epsilon \in \{0.015,0.01,0.0085\}$ over 20 runs. This plot agrees fairly well with the heuristic \eqref{IneqHeuristic}, as do other generated plots. The most important property of our heuristic is that one need not spend an unlimited amount of computational resources to learn a `good enough' partitioning of the state space, and that the computational resources required can be very modest if there exists a very good partition. Similarly to \cite{DLS12}, for problems where we expect the method in this paper to work very well, we find the repartitioning step to be computationally inexpensive.

\bibliographystyle{plainnat}
\bibliography{database}

\end{document}